\documentclass{elsart}

\usepackage{amssymb}

\begin{document}
\begin{frontmatter}

\title{On the thermodynamic stability conditions of Tsallis' entropy}

\author{Wada Tatsuaki\corauthref{cor1}}
\ead{wada@ee.ibaraki.ac.jp}
\address{Department of Electrical and Electronic Engineering, 
Ibaraki University, Hitachi,~Ibaraki, 316-8511, Japan}
\corauth[cor1]{Corresponding author.}

\begin{abstract}
The thermodynamic stability condition (TSC) of Tsallis' entropy is revisited.
As Ramshaw [Phys. Lett. A {\bf 198} (1995) 119] has already pointed out,
the concavity of Tsallis' entropy with respect to the internal energy is not 
sufficient to guarantee thermodynamic stability for all values of $q$ 
due to the non-additivity of Tsallis' entropy. Taking account of 
the non-additivity the differential form of the TSC for Tsallis entropy is 
explicitly derived. 
It is shown that the resultant TSC for Tsallis' entropy is equivalent to
the positivity of the standard specific heat. 
These results are consistent with the relation between Tsallis and
R\'enyi entropies.

\end{abstract}

\end{frontmatter}

Since the pioneering work of Tsallis in 1988 \cite{Tsallis88} there
has been growing interest in the nonextensive thermostatistics
\cite{Tsallis99} based on Tsallis' entropy defined by
\begin{equation}
  S_q^{\rm T} = \frac{1 - \sum_i p_i^q}{q - 1},
\label{Sq}
\end{equation}
where $p_i$ is a probability that a system of interest in state $i$,
and $q$ is a real parameter. For the simplicity the Boltzmann constant
sets to unity throughout this paper. The nonextensive thermostatistics
is a generalization of the standard Boltzmann-Gibbs (BG) statistical
mechanics by the real parameter $q$. In the $q \to 1$ limit, $S_q^{\rm
T}$ reduces to BG entropy $S_1^{\rm T} = -\sum_i p_i \ln p_i$.  One of
the most distinct properties of Tsallis' entropy is the {\it
pseudo-additivity}:
\begin{equation}
  S_q^{\rm T}(A, B) =  S_q^{\rm T}(A) + S_q^{\rm T}(B) + (1-q)
  S_q^{\rm T}(A) S_q^{\rm T}(B),
  \label{pseudo-additivity}
\end{equation}
\noindent where $A$ and $B$ are two subsystems in the
sense that the probabilities of the composed system $A+B$ factorize
into those of $A$ and of $B$. The entropic index $q$ characterizes the
degree of non-extensivity of the system through the pseudo-additivity.

It is shown that some important thermodynamical properties, such as
the Legendre structure \cite{Curado91,Tsallis98,Abe01}, hold true for
the nonextensive thermostatistics. The thermodynamic stability plays
an important role in any formalism related to thermodynamics. The
conventional thermodynamic stability arguments \cite{Callen} are based
on the extremum principle and the additivity of entropy, whereas the
nonextensive thermostatistics is based on the non-additive entropy
$S_q^{\rm T}$.  Therefore the thermodynamic stability in the
nonextensive thermostatistics is one of the most important and
nontrivial properties.  For the earlier versions of the nonextensive
thermostatistics \cite{Tsallis88,Curado91}, Ramshaw \cite{Ramshaw95}
had first indicated that the concavity of $S_q^{\rm T}$ is not
sufficient to guarantee thermodynamic stability of the nonextensive
Tsallis entropy $S_q^{\rm T}$ for all values of $q$. Next Tsallis
\cite{Tsallis95} had shown that if we use the un-normalized internal
energy $U_{q,{\rm un}} \equiv \sum_i p_i^q \epsilon_i$, instead of
standard one $U \equiv \sum_i p_i \epsilon_i$, the concavity (or
convexity) of $S_q^{\rm T}$ with respect to $U_{q,{\rm un}}$, i.e., $q
(\partial^2 S_q^{\rm T} / \partial U_{q, {\rm un}}^2) \le 0$, is
recovered. He thus concluded that the thermodynamic stability for all
$q$ is rescued if $U_{q,{\rm un}}$ is used. For the latest version
\cite{Tsallis99,Tsallis98} of nonextensive thermostatistics, in which
the normalized internal energy $U_q \equiv \sum_i p_i^q \epsilon_i /
\sum_j p_j^q$ is used, de Silva {\it et al.}  \cite{Silva00} shown
that $C/q$ is nonnegative for $q \notin [0,1)$, where $C$ stands
for the specific heat defined by $C \equiv \partial U_q
/ \partial T$ and $T$ is the temperature. 
Lenzi {\it et al.}  \cite{Lenzi01} have shown
the specific heat $C$ is positive for all values of $q$
when the normalized Tsallis' entropy \cite{Rajagopal99} defined by
$S_{q,{\rm no}}^{\rm T} \equiv S_q^{\rm T} / \sum_i p_i^q$, is used
instead of $S_q^{\rm T}$

However, what they have shown are the concavity (or convexity)
of Tsallis' entropy with respect to the appropriate
internal energy or the positivity (or negativity) of the specific heat. 
As Ramshaw has originally pointed out, the concavity
of $S_q^{\rm T}$ with respect to internal energy is not sufficient to
guarantee the thermodynamic stability of nonextensive systems
described by $S_q^{\rm T}$, since the conventional thermodynamic
stability arguments \cite{Callen} are based on the additivity of
entropy and since $S_q^{\rm T}$ is not additive.  The essence of
thermodynamic stabilities lies in the entropy maximum principle
\cite{Callen}.  For an additive entropy $S(U)$, e.g., Boltzmann-Gibbs
or R\'enyi one, the relation between the concavity of $S(U)$ and
thermodynamic stability is straightforward as follows. If we were
remove an amount of energy $\Delta U$ from one of two identical
subsystems and transfer it to the other subsystem, the total entropy
would change from its initial value of $2 S(U)$ to $S(U+\Delta U) +
S(U-\Delta U)$. The entropy maximum principle demands that the
resultant entropy is not larger than the initial entropy, i.e.,
\begin{equation}
   2 S(U) \ge S(U+\Delta U) + S(U-\Delta U),
  \label{TSC}
\end{equation}
which is the TSC for the additive entropy $S(U)$.  In the limit of
$\Delta U \to 0$, Eq. (\ref{TSC}) reduces to its differential form
\begin{equation}
   \frac{\partial^2 S(U)}{\partial U^2} \le 0,
  \label{concavity}
\end{equation}
which states the concavity of $S(U)$.  Thus the TSC and the concavity
of $S(U)$ with respect to $U$ are equivalent each other when $S(U)$ is
additive.  However this is not the case for the non-additive entropy
$S_q^{\rm T}$.  In this letter the TSC for the non-additive Tsallis'
entropy $S_q^{\rm T}$ is reconsidered.

Let us begin by deriving the differential form of the TSC for
$S_q^{\rm T}$.  Taking account of the pseudo-additivity of
Eq. (\ref{pseudo-additivity}), the TSC for $S_q^{\rm T}$ can be
written by
\begin{eqnarray}
  2 S_q^{\rm T}(E) + (1-q) \left[ S_q^{\rm T}(E)\right ]^2 & \ge &
  S_q^{\rm T}(E + \Delta E) + S_q^{\rm T}(E - \Delta E) \nonumber \\
  && + (1-q) S_q^{\rm T}(E + \Delta E) S_q^{\rm T}(E - \Delta E),
  \label{TSC-Sq}
\end{eqnarray}
where $E$ stands for an additive internal energy, e.g., $U$ or $U_q$.
The physical meaning of this inequality is same as that of the
conventional TSC Eq. (\ref{TSC}). If we were remove an amount of
energy $\Delta E$ from one of two identical subsystems and transfer it
to the other subsystem, the total entropy would change from its
initial value of the right-hand-side to the left-hand-side of 
Eq. (\ref{TSC-Sq}).  The
entropy maximum principle demands that the resultant entropy is not
larger than the initial entropy.  The difference between
Eqs. (\ref{TSC-Sq}) and (\ref{TSC}) is the presence of the nonlinear
terms proportional to $1-q$, which originally arise from the
pseudo-additivity of $S_q^{\rm T}$.

For $\Delta E \to 0$, Eq. (\ref{TSC-Sq}) reduces to its differential
form:
\begin{equation}
   \frac{\partial^2 S_q^{\rm T}(E)}{\partial E^2} + (1-q) \left\{
     S_q^{\rm T}(E) \frac{\partial^2 S_q^{\rm T}(E)} {\partial E^2} -
     \left( \frac{\partial S_q^{\rm T}(E)}{\partial E} \right)^2
     \right\} \le 0,
  \label{diff-TSC-Sq}
\end{equation}
Introducing the generalized temperature $T_q$,
\begin{equation}
  \frac{1}{T_q} \equiv \frac{\partial S_q^{\rm T}}{\partial E},
\end{equation}
and generalized specific heat $C_q$,
\begin{equation}
  \frac{1}{C_q} \equiv \frac{\partial T_q}{\partial E} 
      = -T_q^2 \frac{\partial^2 S_q^{\rm T}}{\partial E^2},
\end{equation}
Eq. (\ref{diff-TSC-Sq}) reduces to
\begin{equation}
   \frac{1 + (1-q)S_q^{\rm T}}{C_q} + (1-q) \ge 0,
   \label{TSC-Cq-Sq}
\end{equation}
or by utilizing the definition Eq. (\ref{Sq}) of $S_q^{\rm T}$ and the
relation $\sum_i p_i^q = Z_q^{1-q}$ \cite{Tsallis99,Tsallis98},
Eq. (\ref{TSC-Cq-Sq}) can be cast into the form,
\begin{equation}
   \frac{Z_q^{1-q}}{C_q} + (1-q) \ge 0,
   \label{TSC-Cq}
\end{equation}
where $Z_q$ is the generalized partition function. 
In this way the positivity of $C_q$ only is not
sufficient to guarantee the TSC for the nonextensive thermostatistics
based on $S_q^{\rm T}$. For $1-q \ge 0$, however, the positivity of
$C_q$ guarantees to satisfy the Eq. (\ref{TSC-Cq}) since $Z_q^{1-q} =
\sum_i p_i^q$ is always positive.  
Note that the generalized temperature $T_q$ is not equivalent to the physical
temperature $T$, which is an intensive quantity and obeys the
thermodynamic zeroth law. They are related by \cite{S_Abe01}
\begin{equation}
   T = \{1+(1-q)S_q^{\rm T} \} \cdot T_q.
   \label{T-Tq}
\end{equation}
By differentiating the both sides of Eq.(\ref{T-Tq}) with respect to $E$, 
we obtain
the relation between the specific heat $C \equiv \partial E/\partial T$ and 
the generalized specific heat $C_q$ as
\begin{equation}
   \frac{1}{C} = \frac{1+(1-q)S_q^{\rm T}}{C_q} + 1-q.
\end{equation}
The TSC condition of Eq. (\ref{TSC-Cq-Sq}) for $S_q^{\rm T}$  is thus 
equivalent to the positivity of the specific heat,
\begin{equation}
  C \ge 0,
  \label{TSC-C}
\end{equation}
which is also equivalent to the conventional TSC.

Similar argument can be applied to
the TSC for the normalized Tsallis entropy $S_{q,{\rm no}}^{\rm T}
\equiv S_q^{\rm T} / \sum_i p_i^q$ but one must take account of the
fact that $S_{q,{\rm no}}^{\rm T}$ obeys the modified
pseudo-additivity,
\begin{equation}
  S_{q, \rm no}^{\rm T}(A, B) = S_{q, \rm no}^{\rm T}(A) + 
          S_{q, \rm no}^{\rm T}(B) + (q-1) S_{q, \rm no}^{\rm T}(A) 
          S_{q, \rm no}^{\rm T}(B),
\end{equation}
instead of the pseudo-additivity Eq. (\ref{pseudo-additivity}).

Now let us investigate the above results from the point of the view of 
the relation between Tsallis and R\'enyi entropies.  Ramshaw
\cite{Ramshaw95} had obtained the same result of Eq. (\ref{TSC-Sq}) by
utilizing the relation Eq. (\ref{T-transform}) between the both entropies.  
R\'enyi entropy
\cite{Renyi} is defined by,
\begin{equation}
  S_q^{\rm R} = \frac{\ln \sum_i p_i^q}{1-q},
\end{equation}
which is additive for statistically independent systems, and a concave
functions of the probabilities $p_i$ for $0<q<1$.  Indeed $S_q^{\rm
T}$ and $S_q^{\rm R}$ are closely related by
\begin{equation}
   S_q^{\rm R} = \frac{ \ln [1+(1-q)S_q^{\rm T}]}{1-q}.
   \label{T-transform}
\end{equation}
For $1-q>0$, $S_q^{\rm R}$ is thus monotonically increasing function
of $S_q^{\rm T}$ and vice versa. The extremization of either entropy
subject to the same constraints will produce the same result.  By
differentiating twice the both sides of Eq. (\ref{T-transform}) with
respect to $E$, we see that the concavity of $S_q^{\rm R}(E)$, which
is the TSC for $S_q^{\rm R}(E)$, is equivalent to
Eq. (\ref{diff-TSC-Sq}).

In order to further study the relation between the TSCs for the
both entropies, we focus on the standard internal energy case ($U$),
since there is no study based on R\'enyi entropy with the
(un-)normalized q-average to the author's knowledge.  Let us review the
connection between the concavity (convexity) of an entropy ${\mathcal
S}$ with respect to the probability distributions ${\bf p} = (p_1,
p_2, \dots)$ and the concavity (convexity) of ${\mathcal S}$ with respect 
to the standard internal
energy $U$.  They are related by ${\mathcal S}(U) = {\mathcal S}({\bf
p}(U))$, where ${\bf p}(U)$ is the probability distribution obtained
by extremizing ${\mathcal S}({\bf p})$ subject to the constraint $U
\equiv \sum_i p_i \epsilon_i$, and $p_i$ is a probability of state $i$
whose energy is $\epsilon_i$.  The result \cite{Ramshaw95} is that the
concavity (convexity) of ${\mathcal S}({\bf p})$ implies the concavity
(convexity) of ${\mathcal S}(U)$.  The essence of the proof lies in
the entropy extremum principle and no need to require the additivity
of ${\mathcal S}$. The result thus holds for also non-additive
Tsallis' entropy!  Since $S_q^{\rm T}({\bf p})$ is concave for $q>0$
\cite{Tsallis88,Tsallis99}, $S_q^{\rm T}(U)$ is concave for $q>0$. On
the other hand Since $S_q^{\rm R}({\bf p})$ is concave for $0<q<1$
\cite{Renyi}, $S_q^{\rm R}(U)$ is concave for $0<q<1$.

Now let us consider what is deduced about the TSC for $S_q^{\rm T}$
from only the concavity of $S_q^{\rm T}(U)$, which holds for $q>0$.
The concavity of $S_q^{\rm T}(U)$ can be written by,
\begin{equation}
   2 S_q^{\rm T}(\frac{U_a+U_b}{2}) \ge S_q^{\rm T}(U_a) + S_q^{\rm
   T}(U_b).
   \label{concavity-Sq}
\end{equation}
With the help of the arithmetic-geometric mean inequality (
$\frac{a+b}{2} \ge \sqrt{ab}$ for $a, b>0$), we readily obtain
\begin{equation}
   \frac{S_q^{\rm T}(U_a) + S_q^{\rm T}(U_b)}{2} \ge \sqrt{S_q^{\rm
T}(U_a) S_q^{\rm T}(U_b)},
   \label{AG-mean}
\end{equation}
since $S_q \ge 0$.  Combining the Eqs. (\ref{concavity-Sq}) and
(\ref{AG-mean}), we obtain
\begin{equation}
   \left[ S_q^{\rm T}(\frac{U_a+U_b}{2}) \right]^2 \ge S_q^{\rm
   T}(U_a) S_q^{\rm T}(U_b).
   \label{ineq}
\end{equation}

Multiplying both sides of Eq. (\ref{ineq}) by $(1-q)$ and adding to
both sides of Eq. (\ref{concavity-Sq}), we obtain
\begin{eqnarray}
  2 S_q^{\rm T}(\frac{U_a+U_b}{2})&+& (1-q) \left[ S_q^{\rm
T}(\frac{U_a+U_b}{2})\right ]^2 \nonumber \\ & \ge & S_q^{\rm T}(U_a)
+ S_q^{\rm T}(U_b) + (1-q) S_q^{\rm T}(U_a) S_q^{\rm T}(U_b),
   \label{nonadditive-TSC}
\end{eqnarray}
if $1-q \ge 0$, which is consistent with the condition that the functional
relation of Eq. (\ref{T-transform}) guarantees the monotonic increase of 
$S_q^{\rm R}$ with $S_q^{\rm T}$ and vice versa. 
Since the concavity of $S_q^{\rm
T}(U)$ holds for $q>0$, the resulting TSC of
Eq. (\ref{nonadditive-TSC}) is satisfied for $0 < q \le 1$.  Within
the same range of $q$ the TSC for $S_q^{\rm R}$ also holds since
$S_q^{\rm R}(U)$ is concave for $0 < q \le 1$.  The TSCs for the both
entropies are therefore satisfied for $0 < q \le 1$.  This is
consistent with the equivalence \cite{Ramshaw95,S_Abe01} between
$S_q^{\rm R}$ and $S_q^{\rm T}$ for $0 < q \le 1$.

In summary, the explicit differential form Eq. (\ref{diff-TSC-Sq}) of
the TSC for $S_q^{\rm T}$ is derived by taking account of the
pseudo-additivity of $S_q^{\rm T}$.  Unlike the TSC for conventional
additive entropy, the concavity of $S_q^{\rm T}(U)$ is not equivalent
to the TSC for $S_q^{\rm T}(U)$ due to the pseudo-additivity.  The TSC
for $S_q^{\rm T}(U)$ is thus not simply related to the positivity of
the generalized specific-heat $C_q$, which is equivalent to the
concavity of $S_q^{\rm T}(U)$, but related to the positivity of 
the specific heat $C$. 
These results are consistent with the relation
between Tsallis and R\'enyi entropies for $0<q \le 1$.
The author acknowledges Prof. S.~Abe for useful comments and reading
the manuscript.


\begin{thebibliography}{00}



\bibitem{Tsallis88} C. Tsallis, J. Stat. Phys. {\bf 52} (1988) 479.

\bibitem{Tsallis99} C. Tsallis, Braz. J. Phys. {\bf 29} (1999) 1;
Chaos, Solitons, \& Fractals, {\bf 13} (2002) 371; and references
there in.

\bibitem{Curado91} E. M. F. Curado, C. Tsallis, J. Phys. A {\bf 24}
(1991) L69.

\bibitem{Tsallis98}
C. Tsallis, R. S. Mendes and A. R. Plastino, Physica A {\bf 261}
(1998) 534.

\bibitem{Abe01}
S. Abe, S. Mart\'inez, F. Pennini and A. Plastino, Phys. Lett. A {\bf
281} (2001) 126.

\bibitem{Callen} H. B. Callen, Thermodynamics and an introduction to
thermostatics, 2nd Ed. (Wiley, New York, 1985).

\bibitem{Ramshaw95}
J. D. Ramshaw, Phys. Lett. A. {\bf 198} (1995) 119.

\bibitem{Tsallis95}
C. Tsallis, Phys. Lett. A. {\bf 206} (1995) 389.

\bibitem{Silva00}
L. R. de Silva, E. K. Lenzi, J. S. Andrade Jr., J. Mendes Filho,
Physica A {\bf 275} (2000) 396.

\bibitem{Lenzi01}
E. K. Lenzi, R. S. Mendes, L. R. da Silva, Physica A {\bf 295} (2001)
230-233.

\bibitem{Rajagopal99}
A. K. Rajagopal, S. Abe, Phys. Rev. Lett. {\bf 83} (1999) 1711.

\bibitem{Renyi} A. R\'enyi, Probability theory (North-Holland,
Amsterdam, 1970)

\bibitem{S_Abe01}
S. Abe, Physica A {\bf 300} (2001) 417.

\end{thebibliography}
\end{document}